# Unambiguous identification of monolayer phosphorene by phase-shifting interferometry

Jiong Yang,[1,†] Renjing Xu,[1,†] Jiajie Pei, [1,4,†] Ye Win Myint,[1] Fan Wang,[2] Zhu Wang,[3] Shuang Zhang,[1] Zongfu Yu,[3] and Yuerui Lu[1*]

[1] Research School of Engineering, College of Engineering and Computer Science, the Australian National University, Canberra, ACT, 0200, Australia

[2] Department of Electronic Materials Engineering, Research School of Physics and Engineering, the Australian National University, Canberra, ACT, 0200, Australia

[3] Department of Electrical and Computer Engineering, University of Wisconsin, Madison, Wisconsin 53706, USA

[4] School of Mechanical Engineering, Beijing Institute of Technology, Beijing, 100081, China

[†] These authors contributed equally to this work

[*] To whom correspondence should be addressed: Yuerui Lu (yuerui.lu@anu.edu.au)

**Monolayer phosphorene provides a unique two-dimensional (2D) platform to investigate the fundamental many-body interactions. However, owing to its high instability, unambiguous identification of monolayer phosphorene has been elusive. Consequently, many important fundamental properties, such as exciton dynamics, remain underexplored. We report a rapid, noninvasive, and highly accurate approach based on optical interferometry to determine the layer number of phosphorene, and confirm the results with reliable photoluminescence measurements. Based on the measured optical gap and the calculated electronic energy gap, we determined the exciton binding energy to be ~0.4 eV for the monolayer phosphorene on $SiO_2$/Si substrate, which agrees well with theoretical predictions. Our results open new avenues for exploring fundamental phenomena and novel optoelectronic applications using monolayer phosphorene.**

Phosphorene is a recently developed two-dimensional (2D) material that has attracted tremendous attention owing to its unique anisotropic manner[1-6], layer-dependent direct band gaps[7,8], and quasi-one-dimensional (1D) excitonic nature[9,10], which are all in drastic contrast with the properties of other 2D materials, such as graphene[11] and transition metal dichalcogenide (TMD) semiconductors[12-14]. Monolayer phosphorene has been of particular interest in exploring technological applications and investigating fundamental phenomena, such as 2D quantum confinement and many-body interactions[9,15]. However, such unique 2D materials are unstable in ambient conditions and degrade quickly[8,16]. Particularly, monolayer phosphorene is expected to be much less stable than few-layer phosphorene[16], hence making its identification and characterization extremely challenging. There is a huge controversy on the identification of very few-layer (one or two layers) phosphorene and thus on their properties[16-18]. This controversy was primarily due to the lack of a robust experimental technique to precisely identify the monolayer phosphorene. Consequently, many important fundamental properties of monolayer phosphorene remain elusive. In this study, we propose and implement a rapid, noninvasive, and highly accurate approach to determine the layer number of mono- and few-layer phosphorene by using optical interferometry. The identification is clearly confirmed by the strongly layer-dependent peak energies in the measured photoluminescence (PL) spectra. Our results provide a new platform for the investigation of fundamental many-body interactions and to explore new optoelectronic applications using monolayer phosphorene.

Both atomic force microscopy (AFM) and Raman spectroscopy have been used to reliably determine the sample thickness of TMD semiconductors with monolayer precision[19]. However, these two methods are not reliable for the identification of very-few-layer phosphorene (one or two layers) in ambient conditions. The scanning rate of AFM is slow compared to the fast degradation of very-few-layer phosphorene in ambient conditions and AFM can also introduce potential contaminants that might affect further characterizations on the same sample. Unlike in TMD semiconductors, where Raman mode frequency has a monotonic dependence on the layer number, phosphorene has a non-monotonic dependence owing to the complicated Davydov-related effects[18]. Moreover, the relatively high-power laser used in Raman spectroscopy can significantly damage the phosphorene samples.

To overcome the aforementioned challenges, we propose and implement a rapid, noninvasive, and highly accurate approach to determine the layer number by using optical interferometry (Figure 1). Specifically, we measure the optical path length (OPL) of the light reflected from the phosphorene that was mechanically exfoliated onto a SiO$_2$/Si substrate (275 nm thermal oxide). The OPL is determined from the relation: $OPL_{BP} = -\frac{\lambda}{2\pi}(\phi_{BP} - \phi_{SiO_2})$, where $\lambda$ is the wavelength of the light source and is equal to 535 nm, and $\phi_{BP}$ and $\phi_{SiO_2}$ are the phase shifts of the light reflected from the phosphorene flake and the SiO$_2$/Si substrate (Figure 1c inset), respectively. The direct relationship between the OPL and the layer number is firmly established by a first-principle calculation and experimental calibration, as shown in Figure 1d. Even though the thickness of monolayer phosphorene is less than 1 nm, its OPL is significantly larger than 20 nm owing to the multiple interfacial light reflections (Supplementary Information). That is, the virtual thickness of a phosphorene flake is amplified by more than 20 times in the optical interferometry, making the flakes easily identifiable. In the experiment, phase-shifting interferometry (PSI) is used to measure the OPL by analyzing the digitized interference pattern. In contrast to the highly focused and relatively high-power laser used in Raman system, PSI uses almost non-focused and very low-density light from a light-emitting diode (LED) source to achieve fast imaging (Supplementary Information), which inflicts no damage to the phosphorene samples. The step change of the OPL is ~20 nm for each additional phosphorene layer, as indicated by the red dots in Figure 1d. Considering that the accuracy of the instrument is ~0.1 nm, a step change of 20 nm yields extremely robust identification of the layer number. Statistical OPL values for phosphorene from mono- to six-layer (1L to 6L) were collected and at least five different samples were measured with the PSI system for each layer number. The measured OPL values agree very well with our theoretical calculations (Figure 1d). Recently, we also successfully used PSI to quickly and precisely identify the layer numbers of TMD atomically thin semiconductors[20] (Figure S2).

Subsequent to PSI measurement, the sample was placed into a Linkam THMS 600 chamber, at a temperature of $-10\ °C$ with a slow flow of nitrogen gas to prevent degradation of the sample[8]. The low temperature ($-10\ °C$) is a very crucial factor because it can freeze the moisture in the chamber and significantly delay the sample degradation. Under $-10\ °C$ and nitrogen protection, monolayer phosphorene samples can survive for several hours in the chamber. However, even in a temperature of $-10\ °C$ with nitrogen gas protection, the monolayer phosphorene sample was found damaged when the power of the pulsed laser was higher than 1.15 µW (Figure S3). When the chamber temperature was raised from $-10\ °C$ to room temperature, the monolayer phosphorene was oxidized immediately and the PL signal disappeared. In contrast to monolayer phosphorene, 2L and 3L phosphorene samples can survive for more than fifteen hours under $-10\ °C$ and for several hours when the chamber temperature was raised to room temperature.

Because of the strongly layer-dependent peak energies and the direct band gap nature of phosphorene, we are able to further confirm the layer number identification by measuring their corresponding peak energies of the PL emission (Figure 2). Figure 2a shows the PL spectrum of the monolayer phosphorene sample as indicated in Figure 1a, with a $522\ nm$ pulsed green laser at a laser power of 1.15 µW. The emission peak of the PL spectrum for monolayer phosphorene is at 711 nm, corresponding to a peak energy of 1.75 eV. This PL peak energy value was measured at −10 °C and it is expected not to vary too much at room temperature. Temperature-dependent PL measurements were conducted on 2L and 3L phosphorene samples from 20 °C down to −70 °C; very minor shifts of -0.112 meV/°C and -0.032 meV/°C with temperature were observed for 2L and 3L phosphorene samples, respectively (Figure S4). Assuming a similar low temperature dependence for monolayer phosphorene, its PL peak energy at room temperature would be only ~1–4 meV lower than the measured value at −10 °C. Combining the results of our previous work[8] on few-layer phosphorene (2L to 5L) with the results obtained from our recent samples (1L to 5L), it can be clearly observed that the peak energy of PL emission shows unambiguous layer dependence (Figure 2b). For each layer number, at least three samples were characterized; the measured peak energies for 1L to 5L phosphorene are 1.75 ± 0.04, 1.29 ± 0.03, 0.97 ± 0.02, 0.84 ± 0.02, and 0.80 ± 0.02 eV, respectively.

The peak energy of PL emission, also termed as optical gap ($E_{opt}$), is the difference between the electronic band gap ($E_g$) and the exciton binding energy ($E_b$) (Figure 2b inset). Owing to the strong quantum confinement effect, free-standing monolayer phosphorene is expected to have a large exciton binding energy of ~0.8 eV[9,15], whereas this value is expected to be only ~0.4 eV for monolayer phosphorene on a $SiO_2$/Si substrate because of the increased screening from the substrate[15]. If we use the calculated electronic energy gap of 2.15 eV[16] and the measured optical gap of 1.75 eV in monolayer phosphorene, the exciton binding energy of monolayer phosphorene on $SiO_2$/Si substrate is determined to be ~0.4 eV, which agrees very well with the prediction[15]. The optical gaps in phosphorene increase rapidly with decreasing layer number because of the strong quantum confinement effect and the van der Waals interactions between the neighboring sheets in few-layer phosphorene[9,21]. The layer-dependent optical gaps, as indicated in Figure 2b, agree very well with the theoretical predictions[7,9].

In conclusion, we report a rapid, noninvasive, and highly accurate approach to determine the layer number of mono- and few-layer phosphorene using PSI. The identification is further confirmed by reliable, highly layer-dependent PL peak energies. These two methods provide definite references for future mono- and few-layer phosphorene layer number identification. Based on the measured optical gap and the calculated electronic energy gap, we determined the exciton binding energy to be ~0.4 eV for the monolayer phosphorene on $SiO_2$/Si substrate, which agrees well with theoretical predictions. Our results open new routes for both the investigation of 2D quantum limit in reduced dimensions and development of novel optoelectronic devices.

**Methods**

**Sample preparation and characterization.** Mono- and few-layer phosphorene were mechanically exfoliated from bulk crystals and drily transferred onto $SiO_2$/Si (275 nm thermal oxide) substrates. A phase shifting interferometer (Vecco NT9100) was used to obtain all the OPL values for phosphorene samples. Monolayer phosphorene samples were put into a Linkam THMS 600 chamber and the temperature was set as −10 °C during the PL measurements. A linearly polarized pulse laser (frequency doubled to 522 nm, with 300 fs pulse width and 20.8 MHz repetition rate) was directed to a high numerical aperture (NA = 0.7) objective (Nikon S Plan 60x). PL signal was collected by a grating spectrometer, thereby recording the PL spectrum through a charge coupled device (CCD) (Princeton Instruments, PIXIS).

For few-layer phosphorene (2L to 5L), the PL measurements were conducted using a T64000 micro-Raman system equipped with a InGaAs detector, along with a 532 nm Nd:YAG laser as the excitation source. For all the PL measurements for 2-5L phosphorene samples, the sample was placed into a microscope-compatible chamber with a slow flow of protection nitrogen gas to prevent sample degradation at room temperature. To avoid laser-induced sample damage, all PL spectra from two- to five-layer phosphorene were recorded at low power level of P ~20 µW.

**Numerical Simulation.** Stanford Stratified Structure Solver (S4)[22] was used to calculate the phase delay. The method numerically solves Maxwell's equations in multiple layers of structured materials by expanding the field in the Fourier-space.

**Acknowledgements**

We wish to acknowledge support from the ACT node of the Australian National Fabrication Facility (ANFF). We also thank Professor Chennupati Jagadish, Professor Lan Fu, and Professor Barry Luther-Davies from the Australian National University (ANU) for facility support. We acknowledge financial support from the ANU PhD scholarship, the China Research Council PhD scholarship, the National Science Foundation (USA) (grant number ECCS-1405201), the Australian Research Council (grant number DE140100805), and the ANU Major Equipment Committee.

**Competing Financial Interests**

The authors declare that they have no competing financial interests.

**FIGURE CAPTIONS**

**Figure 1 | Robust identification of mono- and few-layer phosphorene by phase shifting interferometry (PSI). a,** Optical microscope image of a monolayer phosphorene (labeled as "1L"). Inset is the schematic of single-layer phosphorene molecular structure. **b,** PSI image of the dash line box area indicated in (a). **c,** PSI measured optical path length (OPL) values along the dash line indicated in (b). Inset is the schematic plot showing the PSI measured phase shifts of the reflected light from the phosphorene flake ($\phi_{BP}$) and the SiO$_2$/Si substrate ($\phi_{SiO_2}$). **d,** OPL values from simulation and experiment PSI measurements for phosphorene samples from 1L to 6L. For each layer number of phosphorene, at least five different samples were characterized for the statistical measurements. The red dash line is the linear trend for statistical data measured with the PSI system.

**Figure 2 | Photoluminescence (PL) characteristics of mono- and few-layer phosphorene. a,** PL spectrum of the monolayer phosphorene sample in Figure 1a measured with a pulse 522 nm green laser and with a laser power of 1.15 µW. **b,** Evolution of PL peak energy with layer number of phosphorene from experimental PL spectra, showing a rapid increase in peak energy as the layer number decreases. For each layer number, at least three samples were characterized. Note: the PL for the monolayer phosphorene was measured at −10 °C, while others were measured at room temperature. For monolayer phosphorene, the difference between PL peak energies at −10 °C and at room temperature is estimated to be less than 5 meV. Inset: schematic energy diagram showing the electronic band gap ($E_g$), the optical band gap ($E_{opt}$), and the exciton binding energy ($E_b$).

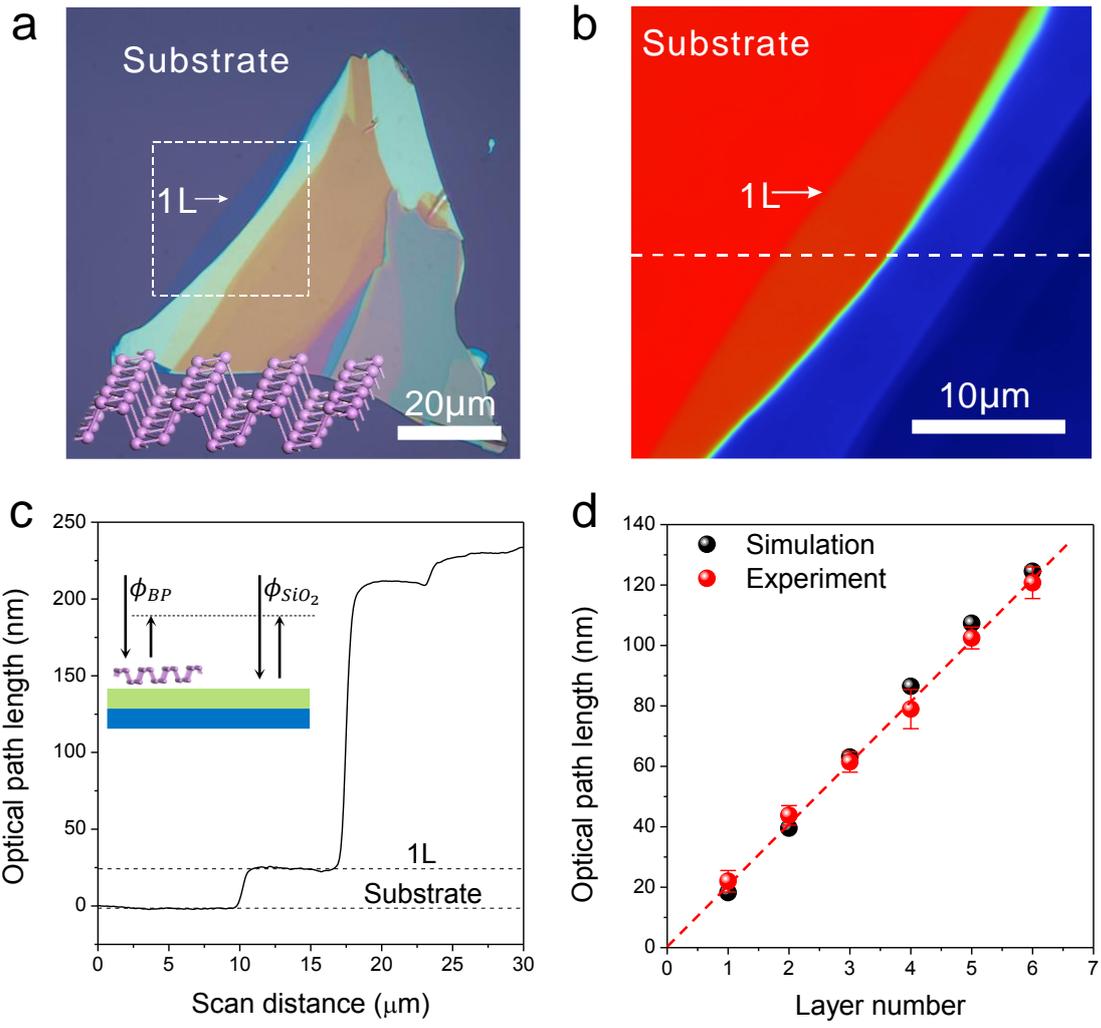

Figure 1

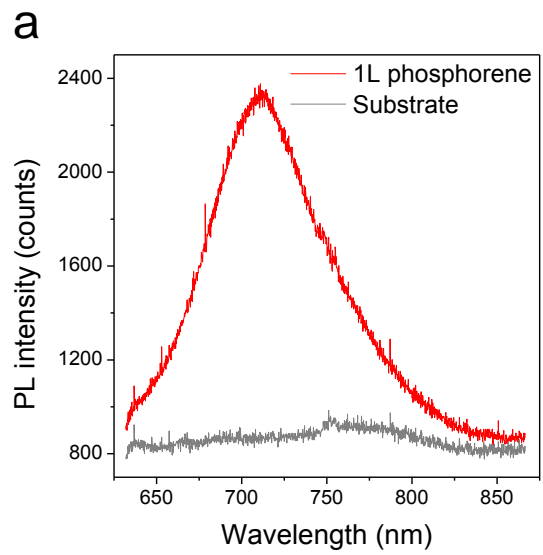 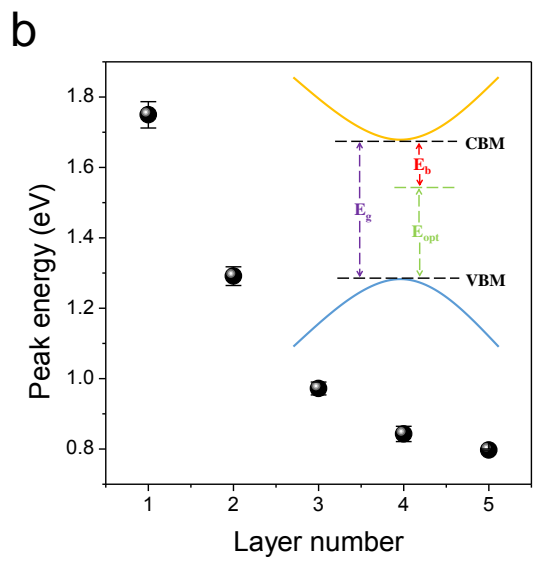

Figure 2

Supplementary Information for

# Unambiguous identification of monolayer phosphorene by phase-shifting interferometry

Jiong Yang,[1,†] Renjing Xu,[1,†] Jiajie Pei,[1,4,†] Ye Win Myint,[1] Fan Wang,[2] Zhu Wang,[3] Shuang Zhang,[1] Zongfu Yu,[3] and Yuerui Lu[1*]

[1]Research School of Engineering, College of Engineering and Computer Science, the Australian National University, Canberra, ACT, 0200, Australia

[2]Department of Electronic Materials Engineering, Research School of Physics and Engineering, the Australian National University, Canberra, ACT, 0200, Australia

[3]Department of Electrical and Computer Engineering, University of Wisconsin, Madison, Wisconsin 53706, USA

[4]School of Mechanical Engineering, Beijing Institute of Technology, Beijing, 100081, China

[†] These authors contributed equally to this work

* To whom correspondence should be addressed: Yuerui Lu (yuerui.lu@anu.edu.au)

## 1. Characterization of another monolayer phosphorene sample

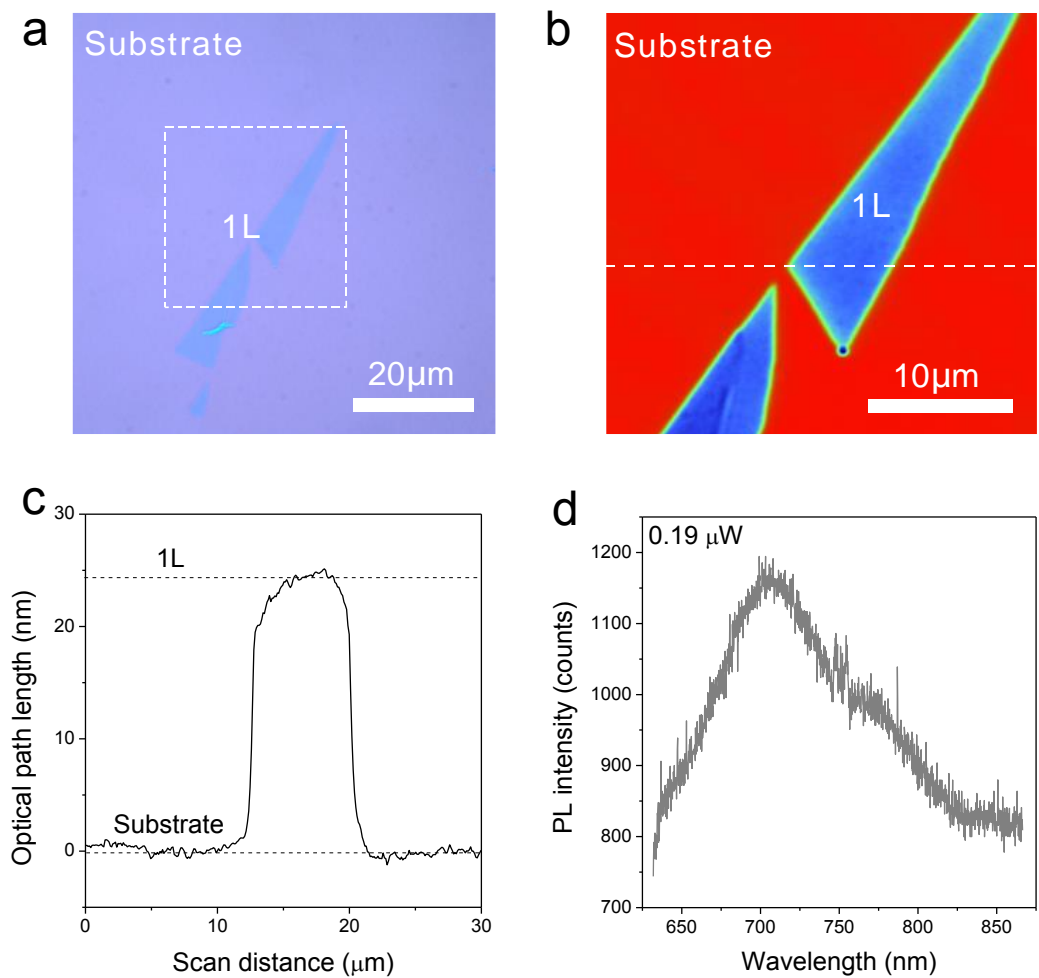

**Figure S1 | Characterization of another monolayer phosphorene. a,** Optical microscope image of monolayer phosphorene (labeled as "1L"). **b,** Phase shifting interferometry (PSI) image of the dash line box area indicated in (a). **c,** PSI measured optical path length (OPL) values along the dash line indicated in (b). **d,** Photoluminescence (PL) spectrum of the monolayer phosphorene sample in (a) measured with a pulse 522 nm green laser at a laser power of 0.19 μW.

## 2. OPL values of mono- and few-layer phosphorene by PSI

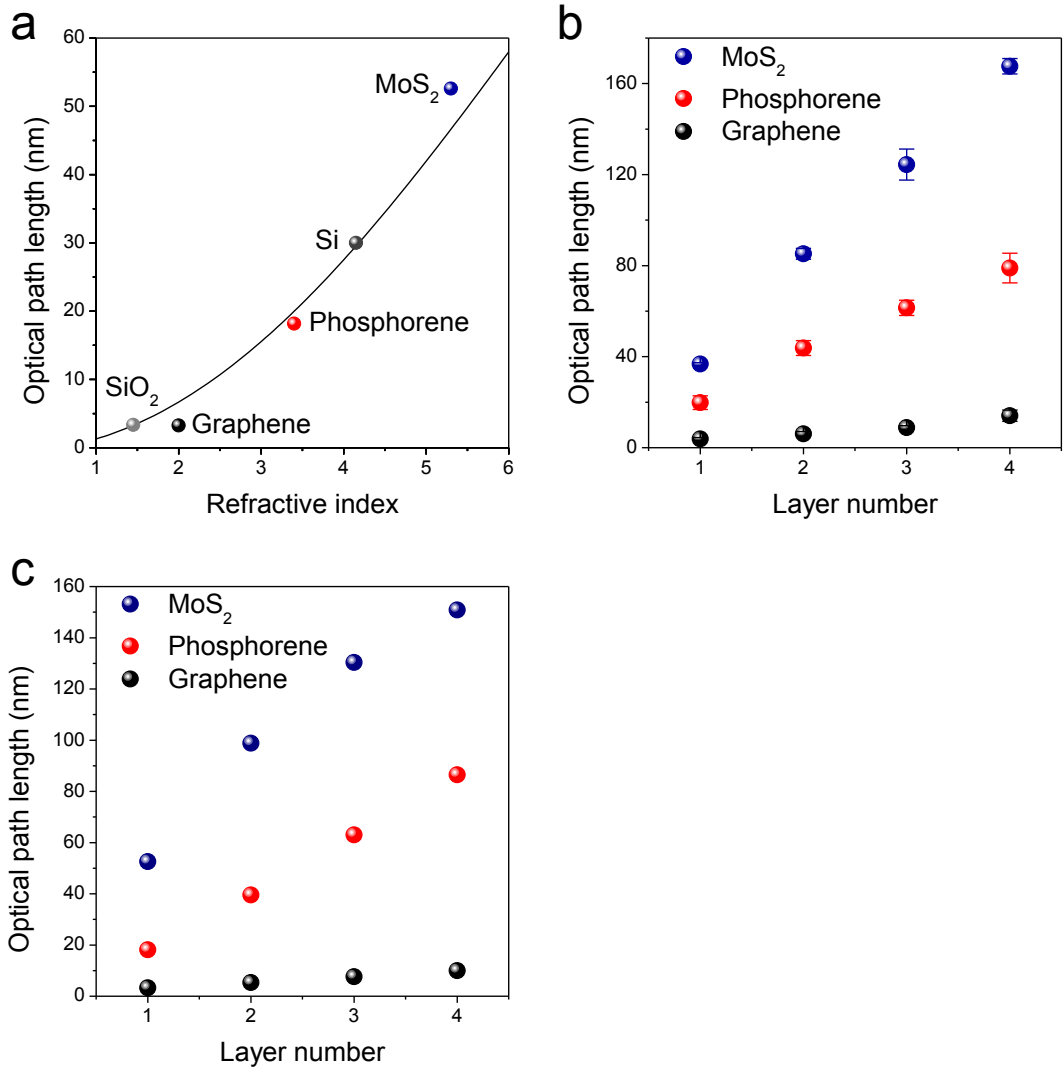

**Figure S2 | OPL values of mono- and few-layer phosphorene in comparison with other materials. a,** Simulated OPL values for light reflected from 2D material (0.67 nm in thickness) with different indices on a $SiO_2$ (275 nm)/Si substrate (solid line). The calculated OPL values of 0.67 nm $SiO_2$, 1L (0.34 nm) graphene, 1L (0.65 nm) phosphorene, 0.67 nm Si and 1L (0.67 nm) $MoS_2$ are represented by markers. **b,** Statistical OPL values of graphene, phosphorene and $MoS_2$ samples from 1L to 4L. For each layer number of these three semiconductors, at least five samples were characterized to get the statistical data, with error bar shown. **c,** Numerical calculation of the OPL values for graphene, phosphorene and $MoS_2$ samples from 1L to 4L.

Numerical calculations were carried out to calculate the OPL values for 2D materials with a thickness of 0.67 nm, shown as solid line in Figure S2a. It is noteworthy that the OPL values increase dramatically with the increase of materials' refractive indices. The OPL values for specific 2D semiconductors, including 0.67 nm $SiO_2$, 1L (0.34 nm) graphene, 1L (0.65 nm) phosphorene, 0.67 nm Si and 1L (0.67 nm) $MoS_2$ are presented in Figure S2a as well. In the numerical calculation, the refractive indices used for $MoS_2$[1], Si, phosphorene[2], graphene[3] and $SiO_2$ were $5.3 + 1.3i$, $4.15 + 0.04i$, $3.4$, $2.6 + 1.3i$ and $1.46$ respectively. Statistical data of OPL values measured with the PSI system and from the numerical calculations for graphene, phosphorene and $MoS_2$ samples from 1L to 4L are presented in Figure S2b and Figure S2c, respectively. Both the measured and calculated OPL values for these three semiconductors show a linear relationship with the layer number and they consist well with each other.

3. Laser induced degradation of monolayer phosphorene

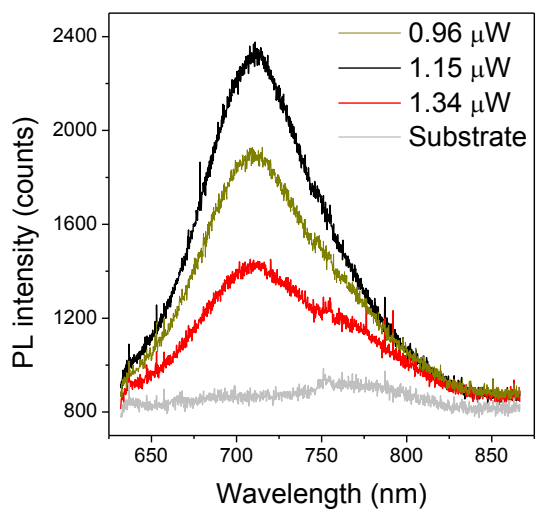

**Figure S3 | Laser induced degradation of monolayer phosphorene.** This sample is the one used for power-dependent PL measurements as shown in Figure 3. Black, dark yellow, red and grey curves represent monolayer phosphorene PL spectra with the laser power of 1.15, 0.96, 1.34 μW and PL spectrum on the $SiO_2$/Si substrate with a laser power of 1.15 μW.

## 4. Temperature-dependent PL peak energies on 2L and 3L phosphorene

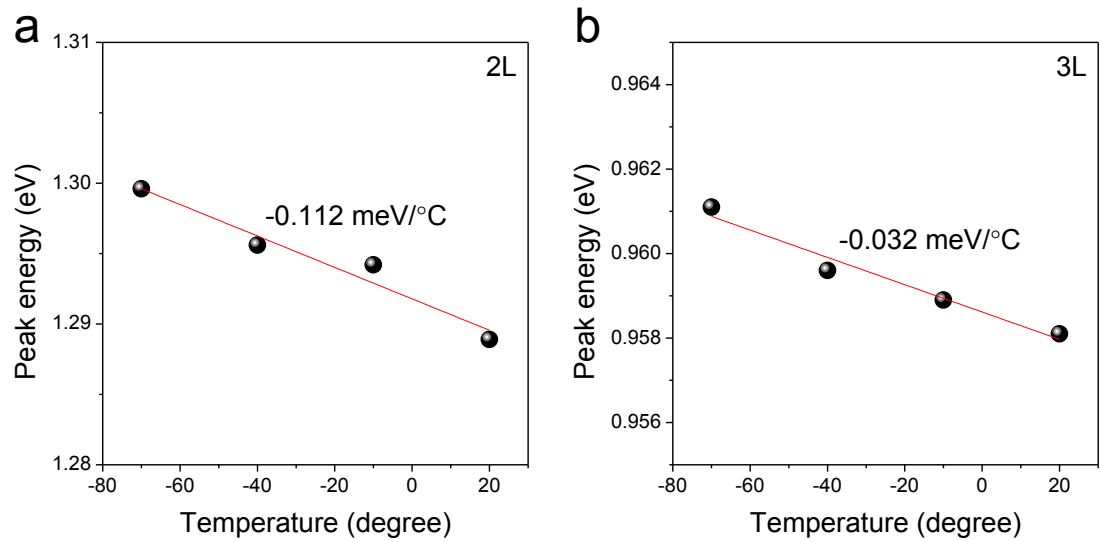

**Figure S4 | Temperature-dependent PL peak positions of 2L (a) and 3L (b) phosphorene samples.**

## 5.     Phase-shifting interferometry (PSI) working principle

PSI was used to investigate the surface topography based on analyzing the digitized interference data obtained during a well-controlled phase shift introduced by the Mirau interferometer[4]. The PSI system (Vecco NT9100) used in our experiments operates with a green LED source centered near 535 nm by a 10 nm band-pass filter[5]. The schematic of the PSI system is shown in Figure S5.

The working principle of the PSI system is as follows[6]. For simplicity, wave front phase will be used for analysis. The expressions for the reference and test wave-fronts in the phase shifting interferometer are:

$$w_r(x,y) = a_r(x,y)e^{i\phi_r(x,y)} \tag{S1}$$

$$w_t(x,y,t) = a_t(x,y)e^{i[\phi_t(x,y)+\delta(t)]} \tag{S2}$$

where $a_r(x,y)$ and $a_t(x,y)$ are the wavefront amplitudes, $\phi_r(x,y)$ and $\phi_t(x,y)$ are the corresponding wavefront phases, and $\delta(t)$ is a time-dependent phase shift introduced by the Mirau interferometer. $\delta(t)$ is the relative phase shift between the reference and the test beam. The interference pattern of these two beams is:

$$w_i(x,y,t) = a_r(x,y)e^{i\phi_r(x,y)} + a_t(x,y)e^{i[\phi_t(x,y)+\delta(t)]} \tag{S3}$$

The interference intensity pattern detected by the detector is:

$$I_i(x,y,t) = w_i^*(x,y,t) * w_i(x,y,t) = I'(x,y) + I''(x,y)\cos[\phi(x,y) + \delta(t)] \tag{S4}$$

where $I'(x,y) = a_r^2(x,y) + a_t^2(x,y)$ is the averaged intensity, $I''(x,y) = 2a_r(x,y) * a_t(x,y)$ is known as intensity modulation and $\phi(x,y)$ is the wavefront phase shift $\phi_r(x,y) - \phi_t(x,y)$.

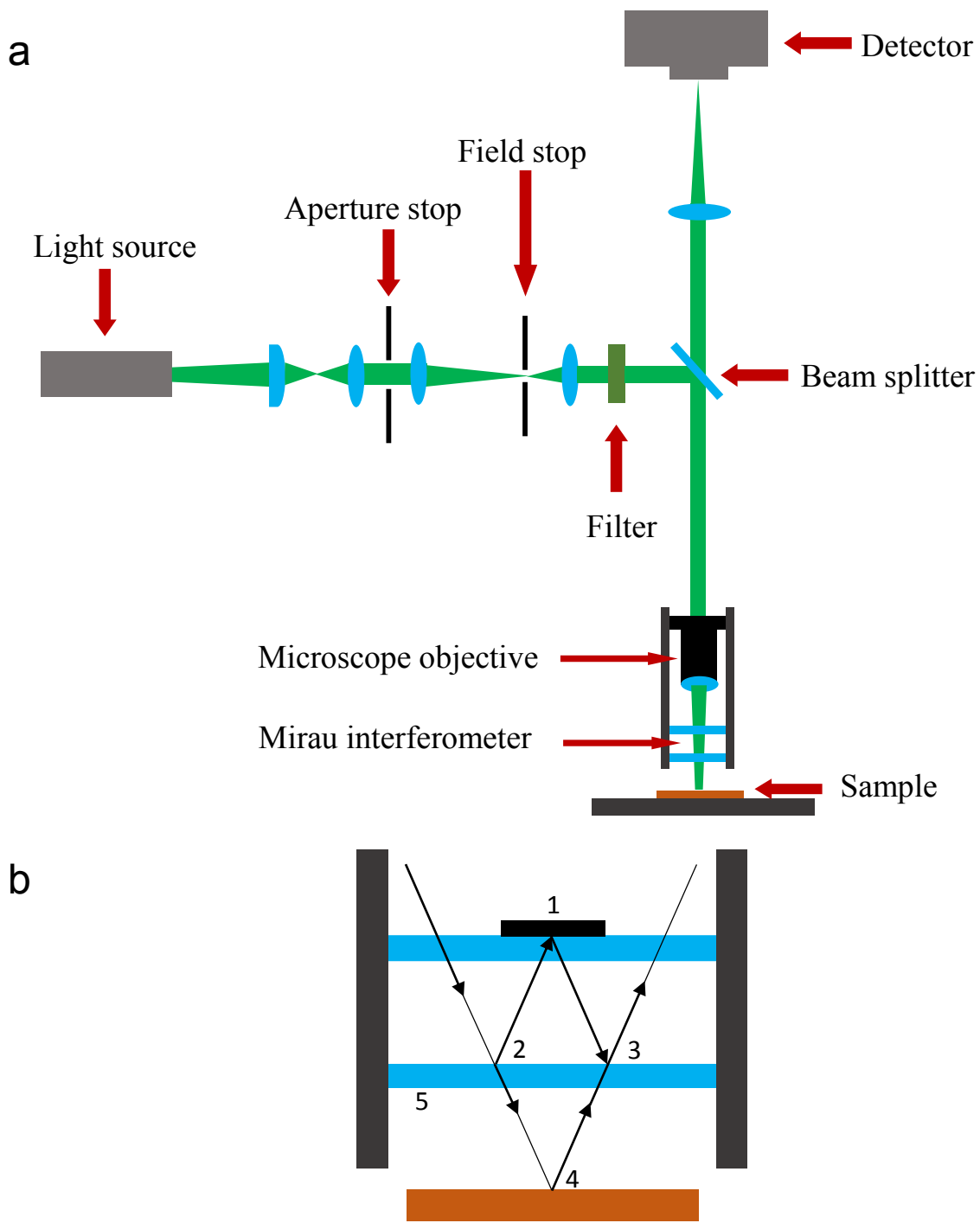

**Figure S5 | Schematic plot of the phase shifting interferometry (PSI) system. a,** Schematic plot of the PSI system. **b,** Zoomed view of the Mirau interferometer. 1. Reference mirror; 2. First reflection of the reference beam; 3. Third reflection of the reference beam; 4. Reflection of the test/objective beam; 5. Semi-transparent mirror. 2-1-3 represents the reference beam and 2-4-3 represents the test/objective beam.

From the above equation, a sinusoidally-varying intensity of the interferogram at a given measurement point as a function of $\delta(t)$ is shown below:

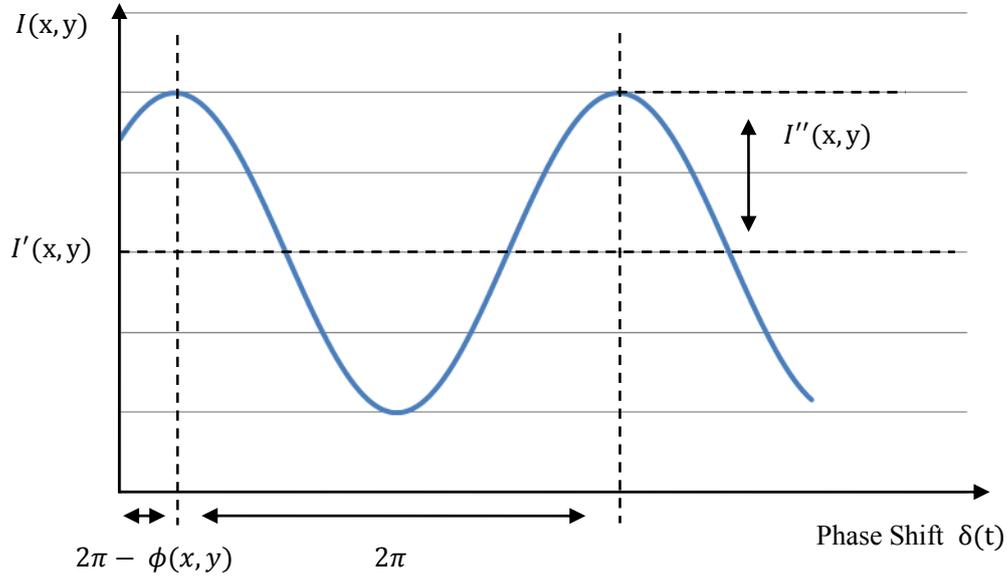

**Figure S6 | Variation of intensity with the reference phase at a point in an interferogram.** $I'(x,y)$ is the averaged intensity, $I''(x,y)$ is half of the peak-to-valley intensity modulation and $\phi(x,y)$ is the temporal phase shift of this sinusoidal variation.

$\delta(t)$ is introduced by the Mirau interferometer, which is shown in Figure S6. When the Mirau interferometer gradually moves toward the sample platform, the optical path length (OPL) of the test beam decreases while the OPL of the reference beam remains invariant.

The computational method of PSI is a four-step algorithm, which needs to acquire four separately recorded and digitalized interferograms of the measurement region. For each separate and sequential recorded interferograms, the phase shift difference is:

$$\delta(t_i) = 0, \frac{\pi}{2}, \pi, \frac{3\pi}{2}; \quad i = 1,2,3,4 \tag{S5}$$

Substituting these four values into the equation S4, leads to the following four equations describing the four measured intensity patterns of the interferogram:

$$I_1(x,y) = I'(x,y) + I''(x,y)\cos[\phi(x,y)] \tag{S6}$$

$$I_2(x,y) = I'(x,y) + I''(x,y)\cos[\phi(x,y) + \frac{\pi}{2}] \tag{S7}$$

$$I_3(x,y) = I'(x,y) + I''(x,y)\cos[\phi(x,y) + \pi] \tag{S8}$$

$$I_4(x,y) = I'(x,y) + I''(x,y)\cos[\phi(x,y) + \frac{3\pi}{2}] \tag{S9}$$

After the trigonometric identity, this yields:

$$I_1(x,y) = I'(x,y) + I''(x,y)\cos[\phi(x,y)] \tag{S10}$$

$$I_2(x,y) = I'(x,y) - I''(x,y)\sin[\phi(x,y)] \tag{S11}$$

$$I_3(x,y) = I'(x,y) - I''(x,y)\cos[\phi(x,y)] \tag{S12}$$

$$I_4(x,y) = I'(x,y) + I''(x,y)\sin[\phi(x,y)] \tag{S13}$$

The unknown variables $I'(x,y)$, $I''(x,y)$ and $\phi(x,y)$ can be solved by only using three of the four equations; but for computational convenience, four equations are used here. Subtracting equation S11 from equation S13, we have:

$$I_4(x,y) - I_2(x,y) = 2I''(x,y)\sin[\phi(x,y)] \tag{S14}$$

And subtract equation S12 from equation S10, we get:

$$I_1(x,y) - I_3(x,y) = 2I''(x,y)\cos[\phi(x,y)] \tag{S15}$$

Taking the ratio of equation S14 and equation S15, the intensity modulation $I''(x,y)$ will be eliminated as following:

$$\frac{I_4(x,y) - I_2(x,y)}{I_1(x,y) - I_3(x,y)} = \tan[\phi(x,y)] \tag{S16}$$

Rearranging equation S16 to get the wave-front phase shift term $\phi(x,y)$:

$$\phi(x,y) = \tan^{-1}\frac{I_4(x,y) - I_2(x,y)}{I_1(x,y) - I_3(x,y)} \tag{S17}$$

This equation is performed at each measurement point to acquire a map of the measured wave-front. Also, in PSI, the phase shift is transferred to the surface height or the optical path

difference (OPD):

$$h(x,y) = \frac{\lambda \phi(x,y)}{4\pi} \qquad (S18)$$

$$OPD(x,y) = \frac{\lambda \phi(x,y)}{2\pi} \qquad (S19)$$

Here, the OPL of the phosphorene flake $OPL_{BP}$ is calculated as:

$$OPL_{BP} = -(OPD_{BP} - OPD_{SiO_2}) = -\frac{\lambda}{2\pi}(\phi_{BP} - \phi_{SiO_2}) \qquad (S20)$$

where $\lambda$ is the wavelength of the light source, $\phi_{BP}$ and $\phi_{SiO_2}$ are the measured phase shifts of the reflected light from the phosphorene flake and the SiO$_2$ substrate, respectively. In our experiments, $\phi_{SiO_2}$ was typically set to be zero, as shown in Figure 1c.

## 6. Calculations for the optical path length (OPL) of atomically thin 2D materials

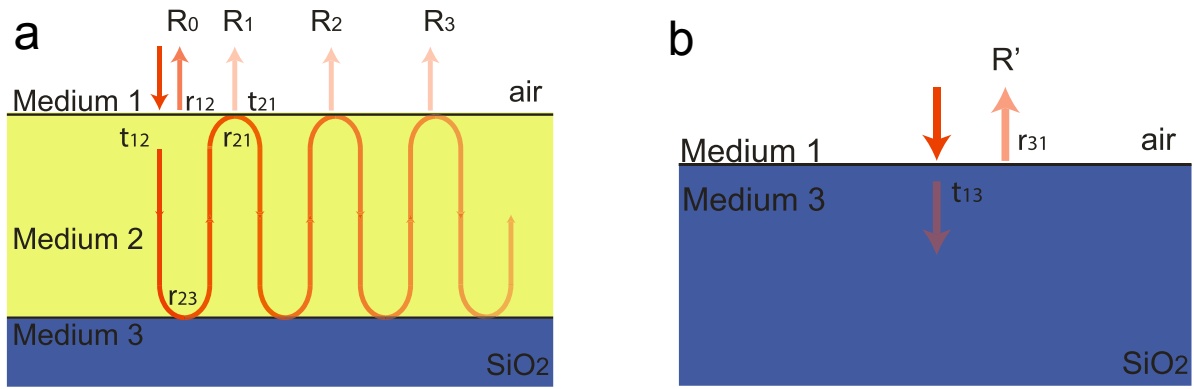

**Figure S7 | a**, Reflection of a three-layer structure. Medium 1 is air, Medium 2 is the 2D material and Medium 3 is an infinite SiO$_2$ substrate. **b**, The reference configuration. Light is incident from air into infinite SiO$_2$ substrate.

The incident light comes from the air resonates inside the 2D material. The total reflection is determined by the interference of all reflected beams $R_i$. To calculate the amplitude of the total reflection, we use $r_{ij}$ ($i,j=1,2,3$) to represent the reflection coefficients when light goes from medium $i$ to medium $j$.

$$r_{ij} = \frac{n_i - n_j}{n_i + n_j} \tag{S21}$$

We use $t_{ij}$ ($i,j$ =1,2,3) to represent the transmission from medium $i$ to medium $j$

$$t_{ij} = \frac{2n_i}{n_i + n_j} \tag{S22}$$

where $n_i$, $n_j$ ($i,j$ =1,2,3) is the refractive index of medium $i,j$. Assuming that the thickness of the 2D material is $d$ and wave vector of incident light in air is $k_0$, we can calculate the reflection of each order,

$$R_0 = r_{12}$$
$$R_1 = t_{12} r_{23} t_{21} e^{i2k_0 nd}$$
$$R_2 = t_{12} r_{23} r_{21} r_{23} t_{21} (e^{i2k_0 nd})^2$$
$$R_3 = t_{12} r_{23} r_{21} r_{23} r_{21} r_{23} t_{21} (e^{i2k_0 nd})^3 \tag{S23}$$

where $2k_0 nd$ is the round trip propagation phase and $n$ is the refractive index of the 2D material. Then the total reflected amplitude is the summation of all reflections, which is

$$\begin{aligned} R &= R_0 + R_1 + R_2 + \\ &= r_{12} + t_{12} r_{23} t_{21} e^{i2k_0 nd} \left[ 1 + r_{21} r_{23} e^{i2k_0 nd} + \left(r_{21} r_{23} e^{i2k_0 nd}\right)^2 + \cdots \right] \\ &= r_{12} + \frac{t_{12} r_{23} t_{21} e^{i2k_0 nd}}{1 - r_{21} r_{23} e^{i2k_0 nd}} \\ &= \frac{1-n}{1+n} + \frac{4n}{(1+n)^2} \frac{(n-1.46)}{(n+1.46)} e^{i2k_0 nd} \frac{1}{1 - \frac{(n-1)(n-1.46)}{(n+1)(n+1.46)} e^{i2k_0 nd}} \end{aligned} \tag{S24}$$

Here we used refractive indices of air and $SiO_2$ as 1 and 1.46, respectively.

The OPL was calculated by comparing the phase difference of the reflected light with and without the 2D material. Figure S7b shows the reference setup. Light is incident directly from air into infinite $SiO_2$ substrate. In this case the reflected amplitude is

$$R' = \frac{n_1 - n_3}{n_1 + n_3} \tag{S25}$$

So we get:

$$OPL = -\frac{\bigl(phase(R)-phase(R')\bigr)}{2\pi}\lambda \tag{S26}$$

where $\lambda$ is the wavelength of light. For phosphorene OPL calculations, we used the measured refractive index from bulk black phosphorus crystals ($n$ = 3.4)[2].

**7. Images and characterization of phosphorene flakes by PSI.**

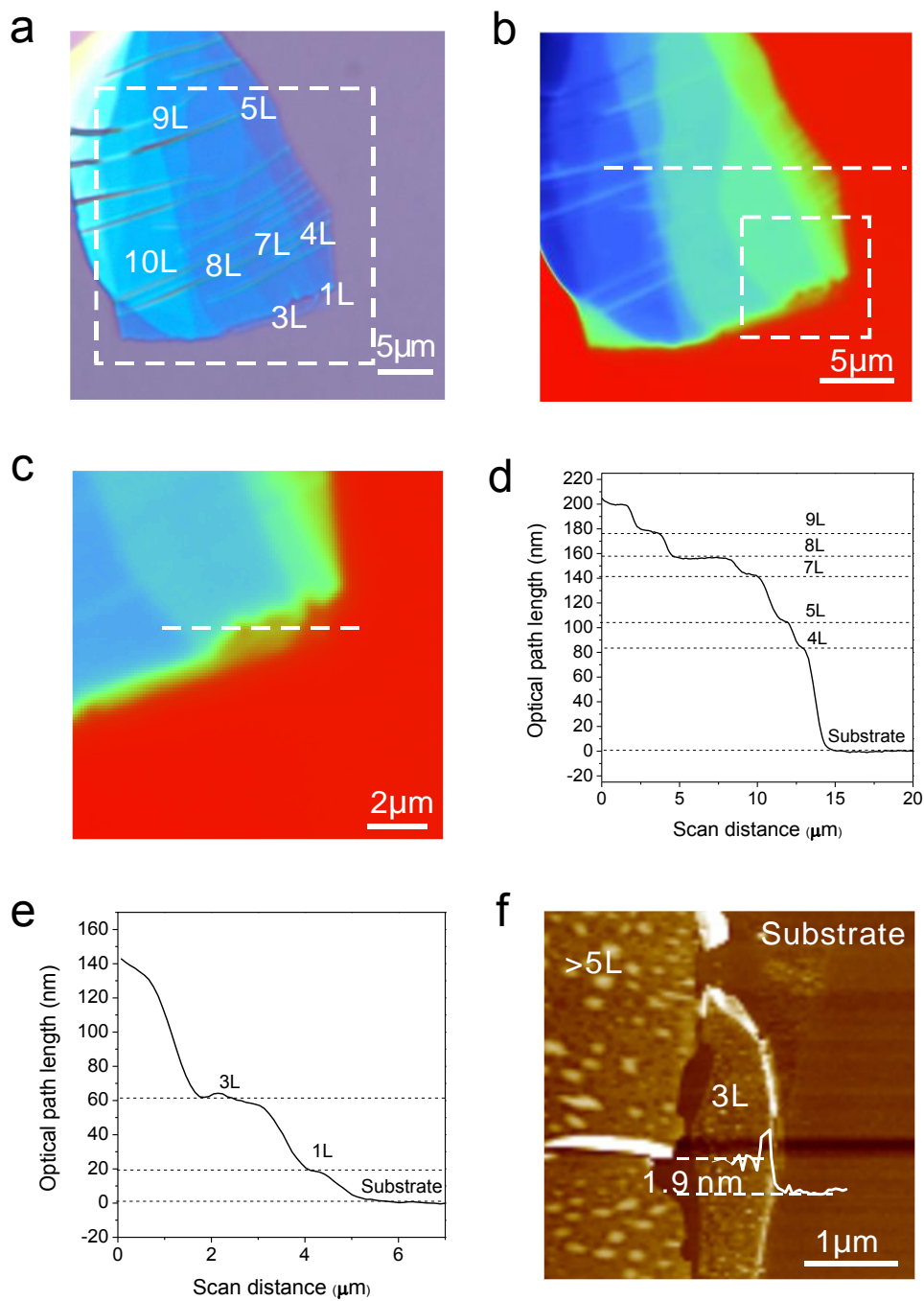

**Figure S8 | Images and characterization of exfoliated phosphorene. a,** Optical microscope image of a phosphorene flake containing multiple layers. **b,** PSI image of the phosphorene flake from the dash line box area indicated in (a). **c,** PSI image of the phosphorene flake from the dash line box area indicated in (b). **d** and **e** display the OPL measured by PSI versus position along the dash line in (b) and (c) respectively. **f,** AFM image of the 3Lphosphorene flake.

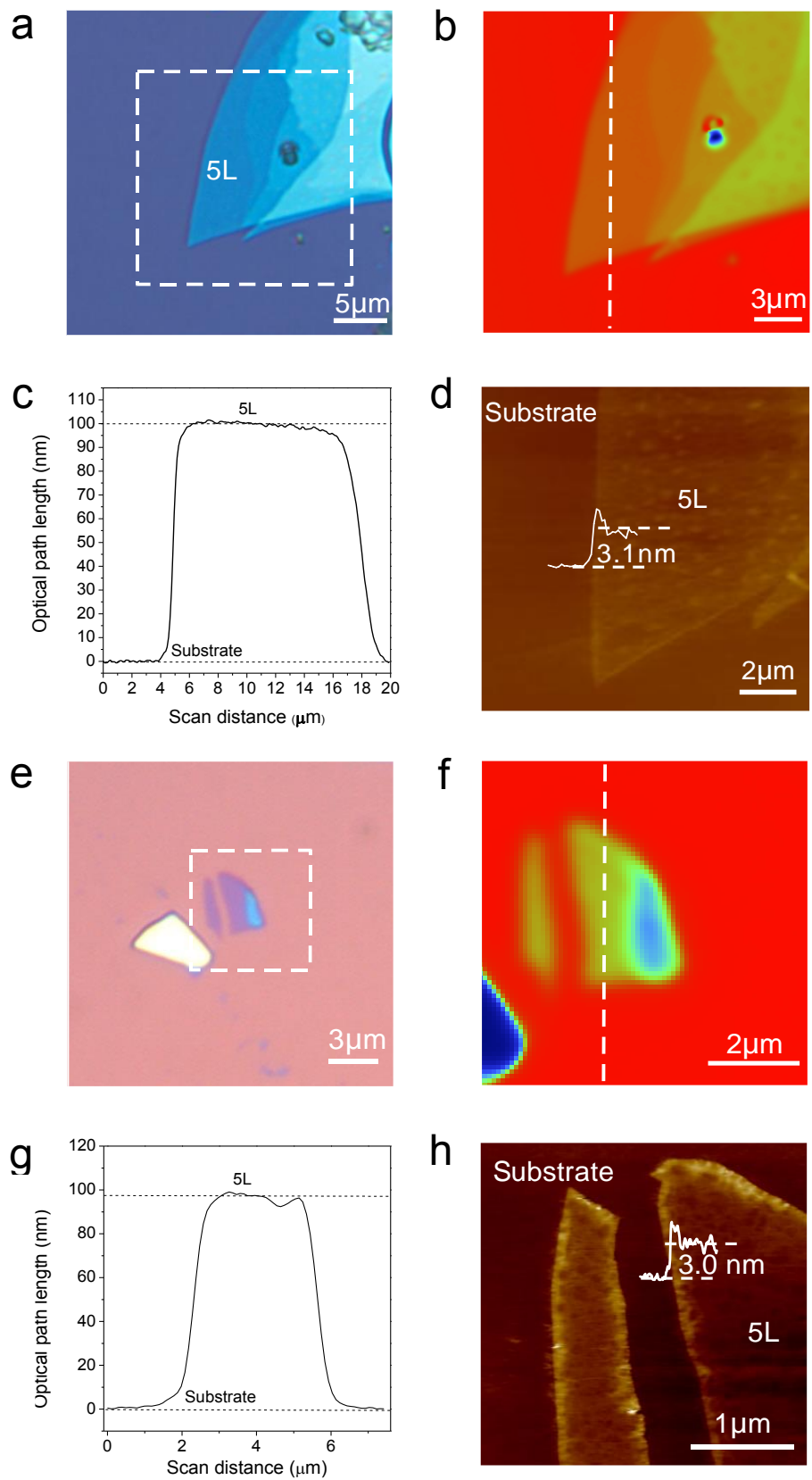

**Figure S9 | Images and characterization of exfoliated 5L phosphorene. a,** Optical microscope image of a 5L phosphorene flake. **b,** PSI image of the 5L phosphorene from the

dash line box area indicated in (a). **c,** OPL measured by PSI along the dash line in (b). **d,** AFM

image of 5L phosphorene. **e, f, g** and **h** display optical microscope image, PSI image (from the

dash line box area indicated in e), OPL (along the dash line in f) and AFM image of another 5L

phosphorene flake, respectively.

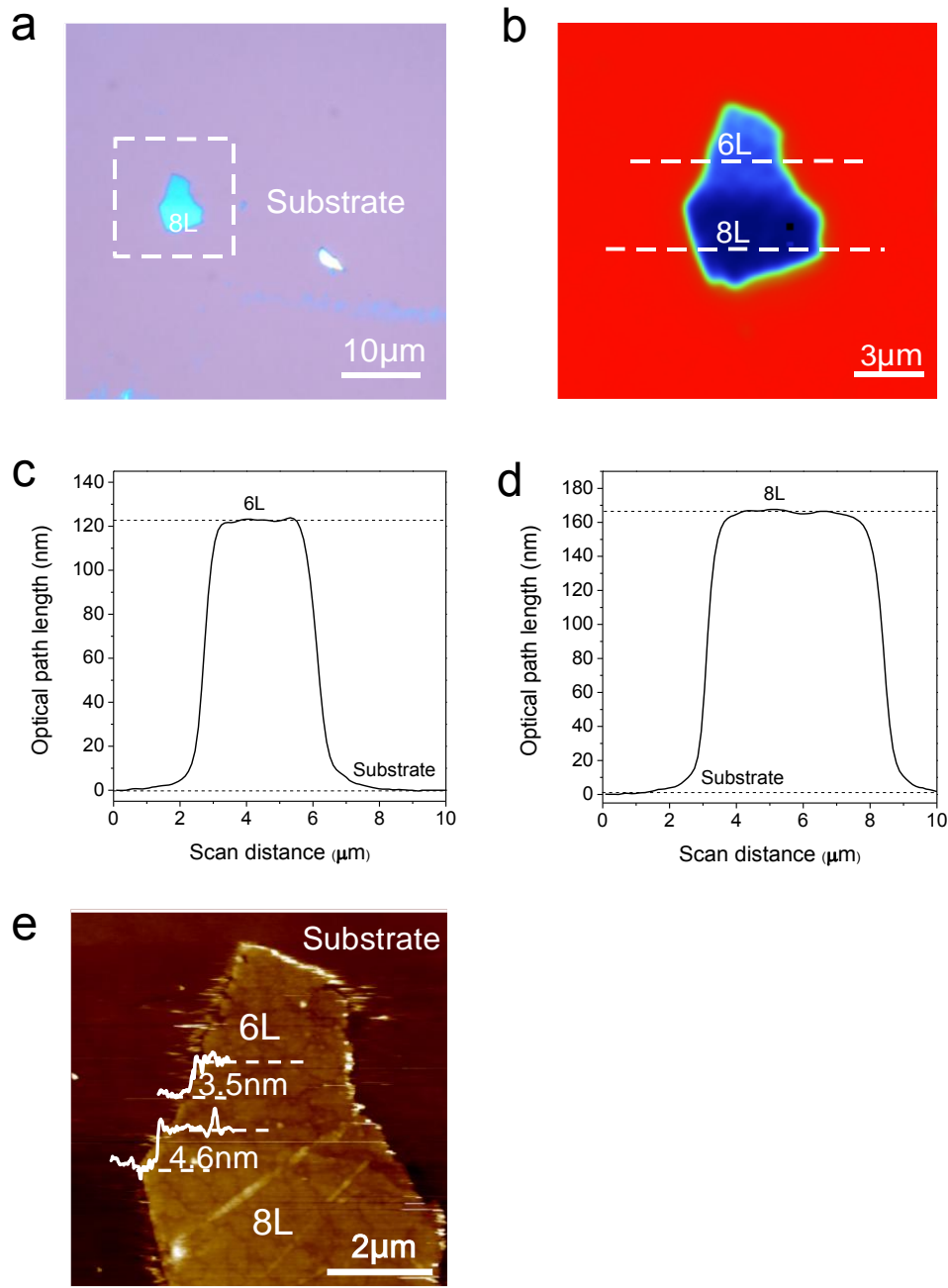

**Figure S10 | Images and characterization of exfoliated 6L and 8L phosphorene flakes. a,**

Optical microscope image of a phosphorene flake containing 6L and 8L. **b,** PSI image of the phosphorene flake from the dash line box area indicated in (a). **c** and **d** display OPL values measured by PSI from the 6L and 8L phosphorene along the dash lines in (b). **e,** AFM image of the 6L and 8L phosphorene.